\documentclass{article}

      \usepackage[preprint]{neurips_2020}

\usepackage[utf8]{inputenc} 
\usepackage[T1]{fontenc}    
\usepackage{hyperref}       
\usepackage{url}            
\usepackage{booktabs}       
\usepackage{amsfonts}       
\usepackage{nicefrac}       
\usepackage{microtype}      
\usepackage{graphicx}       

\title{A Multi-agent Reinforcement Learning Study of Evolution of Communication and Teaching under Libertarian and Utilitarian Governing Systems}

\author{%
  Aslan S.~Dizaji\thanks{The code of this project is available in \url{https://github.com/aslansd/modified-ai-economist-wc} (The Modified AI-Economist with Communication) and \url{https://github.com/aslansd/modified-ai-economist-wt} (The Modified AI-Economist with Teaching).} \\
  AutocurriculaLab\\
  Tehran, Iran\\
  \texttt{asataryd@umich.edu} \\
}

\begin{document}

\maketitle

\begin{abstract}
Laboratory experiments have shown that communication plays an important role in solving social dilemmas. Here, by extending the AI-Economist, a mixed motive multi-agent reinforcement learning environment, I intend to find an answer to the following descriptive question: which governing system does facilitate the emergence and evolution of communication and teaching among agents? To answer this question, the AI-Economist is extended by a voting mechanism to simulate three different governing systems across individualistic-collectivistic axis, from Full-Libertarian to Full-Utilitarian governing systems. In the original framework of the AI-Economist, agents are able to build houses individually by collecting material resources from their environment. Here, the AI-Economist is further extended to include communication with possible misalignment \textendash a variant of signaling game \textendash by letting agents to build houses together if they are able to name mutually complement material resources by the same letter. Moreover, another extension is made to the AI-Economist to include teaching with possible misalignment \textendash again a variant of signaling game \textendash by letting half the agents as teachers who know how to use mutually complement material resources to build houses but are not capable of building actual houses, and the other half as students who do not have this information but are able to actually build those houses if teachers teach them. I found a strong evidence that collectivistic environment such as Full-Utilitarian system is more favourable for the emergence of communication and teaching, or more precisely, evolution of language alignment. Moreover, I found some evidence that evolution of language alignment through communication and teaching under collectivistic governing systems makes individuals more advantageously inequity averse. As a result, there is a positive correlation between evolution of language alignment and equality in the society.
\end{abstract}

\section{Introduction}
Multi-agent reinforcement learning (MARL) comprised of multiple decision making units each one is able to observe the state of and act upon environment to achive its goal. An agent changes the state of the environment by taking an action and then receives a reward and a new observation. This loop continues until the agent achieves its goal or the time-step of the environment reaches to its maximum limit. Each MARL problem has multiple dimensions. First, what are the number of agents, the number of states in the environment, and the number of possible actions of the agents. Second, what kind of knowledge the agents have about the environment, e.g. do they know the state transition probabilities of the environment. Third, what is the scope of their observations, e.g. can they observe the full or partial state of the environment, or the actions and rewards of other agents. Fourth, do the agents operate in zero-sum, general-sum, or common reward situations. Fifth, what kind of objectives the agents have, e.g. what kind of equilibrium they want to reach. Sixth, how much their training and execution are centralized or if there is any communication among them. Moreover, each MARL problem faces at least four challenges. These include the non-stationary caused by learning of multiple interacting agents, unknown optimality of the final selected joint policy or equilibrium, multi-agent credit assignment, and finally the challenge of scaling to a large number of agents (\cite{Albrecht2023}). Furthermore, in the case of mixed motive games, we have two additional challenges of heterogeneous incentives and the difficulty of defining a suitable collective reward function (\cite{Du2023}). Additionally, each MARL problem can be framed in one of three agendas: computational agenda in which the goal of MARL is to compute the solutions for game models, the prescriptive agenda in which the focus is on behavior and performance of the agents during learning, and the descriptive agenda in which the goal of MARL is to simulate the actual behaviour of a population of humans or animals (\cite{Albrecht2023}).

The AI-Economist is a two-level MARL framework (\cite{Zheng2022}), comprised of one single agent as a rational social planner who designs a particular mechanism or policy generally having a goal of optimizing a particular kind of social welfare functions in the society. The other agents are a set of rational economic mobile agents who behave in response to the implemented mechanism or policy generally following their own self-interest. This framework has been used to model the tax-gaming behaviour of agents - optimizing their labors, trading, and building, while the central social planner maximizes productivity or equality in the society (\cite{Zheng2022}). More preciseley, the agents in the Gather-Trade-Build environment of the AI-Economist make efforts to move, gather wood and stone from the environment, trade them with each other via double-auctions using coins as a mean of exchange, and finally \textendash contingent on their build-skill \textendash build houses to earn incomes. On the other hand, the social planner aims to find an optimized taxing schedule to increase productivity or equality in the society (\cite{Zheng2022}). As it is clear, the agenda of this framework is descriptive. The number of mobile agents is between 2 and 10 which is a reasonable choice in MARL. The game is a mixed motive partially observable stochastic game with simultaneous cooperation and competition. The agents share their weights in a centralized training and decentralized execution by having their own set of observations. The non-stationary of the learning agents is partially overcome by curriculum learning and entropy regularization, while the optimality of the selected equilibrium is partially confirmed by letting the environment to go through a very large number of time-steps. Due to complexity of the environment, a two-level Proximal Policy Optimization (PPO) gradient method as a deep reinforcement learning technique is used to solve the equations. Additionally, in the original framework of the AI-Economist, the agents do not have any communication capabilities. Previously, I extended the AI-Economist \textendash called the Modified AI-Economist \textendash in two different directions (\cite{Dizaji2023a, Dizaji2023b}). In the first paper (\cite{Dizaji2023a}), I investigated the impacts of the governing systems or institutions on the origin of morality, prosperity or equality, and fairness in the society. Particularly, I could show that along individualistic-collectivistic axis, the libertarian government generates more fair and moral individuals. Simultaneously, the prosperity is higher in the individualistic libertarian government while the equality is lower. Finally, I could not find any significant difference between the fairness level of the individualistic versus collectivistic governments. In the second paper (\cite{Dizaji2023b}), considering two parallel environments in which one of them is comprised of band-like isolated and the other one of uniformly distributed natural resources, I could show that if the central planner is an arbitrary ruler, each environment evolves through a different path. Band-like environment finally converges to an environment in which all the agents are getting powerless in front of the naked power of the arbitrary governance, while the central planner's net total tax revenue is also getting zero. On the other hand, the uniform environment converges to a final situation in which the society is getting composed of stratified distinct social classes, and the central planner is also able to continue collecting the non-zero taxes. Overall, these two papes are another manifestation of the power of multi-agent reinforcement learning to model social and economical phenomena (\cite{Trott2021, Zheng2022, Zhang2022, Leibo2019, Leibo2021, Johanson2022}).

Here, while keeping the previous extension of the AI-Economist regarding the modelling of governing system along individualistic-collectivistic axis \textendash from Full-Libertarian to Full-Utilitarian governing systems \textendash through a voting mechanism, I further extend the AI-Economist in another direction. The main question here is about which governing system is more favourable for the evolution of communication and teaching through language alignment. Previously, in MARL literature, it has been shown that communication can enhance exploration, maximize reward, and diversify solutions in complex optimisation simulations (\cite{Du2023}). Also, in human experiments, it has been shown that costly punishment has positive effects on resolving common pool resource dilemmas only when it is combined with communication (\cite{Janssen2010}). Basically, human experiments show that the essence of communication even without any enforcement is more effective than the content of the communication in facilitating cooperation in social dilemmas (\cite{Hertz2023}). Several possible mechanisms have been proposed including communication might allow coordination among participants, thus they might develop trust relationships, and communication might express social pressure (\cite{Hertz2023}). Here, in this project, communication and teaching are modelled through two simple variants of signaling game, a game which was originally devised to simulate the emergence of convention (\cite{Karch2023, Ohmer2022}). In game theoretic words, a convention is a system of arbitrary rules that enables two players to share meaningful information. In the Modified AI-Economist with Communication, it is assumed that the agents beside being able to build two types of houses individually, they are able to build the same types of houses with another agent if they communicate with each other and name the two required complementary resources of a house type with the same letters. If their language is aligned they build a house together and obtain a large reward. Otherwise, they align their language gradually and obtain small amount of reward until their language is fully aligned to build houses or an episode is ended. In the Modified AI-Economist with Teaching, the agents are divided to two groups, teachers and students. The teachers know how to combine two complementary natural resources to build a house type but they are not capable of, while the students do not know which complementary natural resources to use to build a house, but if a teacher tell them, they are capable of building. Again, here the communication between teachers and students can be misaligned, i.e. they use different letters for the same natural resources. However, through many interactions between teachers and students, they gradually align their communication \textendash simultaneously gather small rewards \textendash until that point that all students use the same language as teachers and to be able to build houses \textendash together with teachers  \textendash and obtain large rewards.

The results show that collectivistic environment such as Full-Utilitarian system is more favourable for the emergence of communication and teaching, or more precisely, evolution of language alignment. Moreover, there is some evidence that evolution of language alignment through communication and teaching under collectivistic governing systems makes individuals more advantageously inequity averse. As a result, there is a positive correlation between evolution of language alignment and equality in a society. Overall, the results of this paper are tentative and should be elaborated more with further delicate investigation in future projects.

\section{The Modified AI-Economist with Communication/Teaching}
For a complete description of the AI-Economist, please refer to Appendix A. Here, three major modifications that are made to the original framework are described.

First, two new resource materials \textendash iron and soil \textendash are added to the environment, and now with four building materials, the number of possible house types is diversified to two: a red house is exclusively can be built from wood and stone, and a blue house is exclusively can be built from iron and soil. Also, each one of these house types can be built via two means: individually by each single agent, or together with two agents. Moreover, there are two kinds of build skills for each agent: the required skill for building alone and the required skill for building together which is higher than the former. Furthemore, the required labour for building a house type alone or together is also different: the latter is higher than the former. However, move and gather skill and labor, and trade labor will be equal and fixed across all materials and agents.

\begin{figure}
	\centering
	\includegraphics[width=0.7\linewidth]{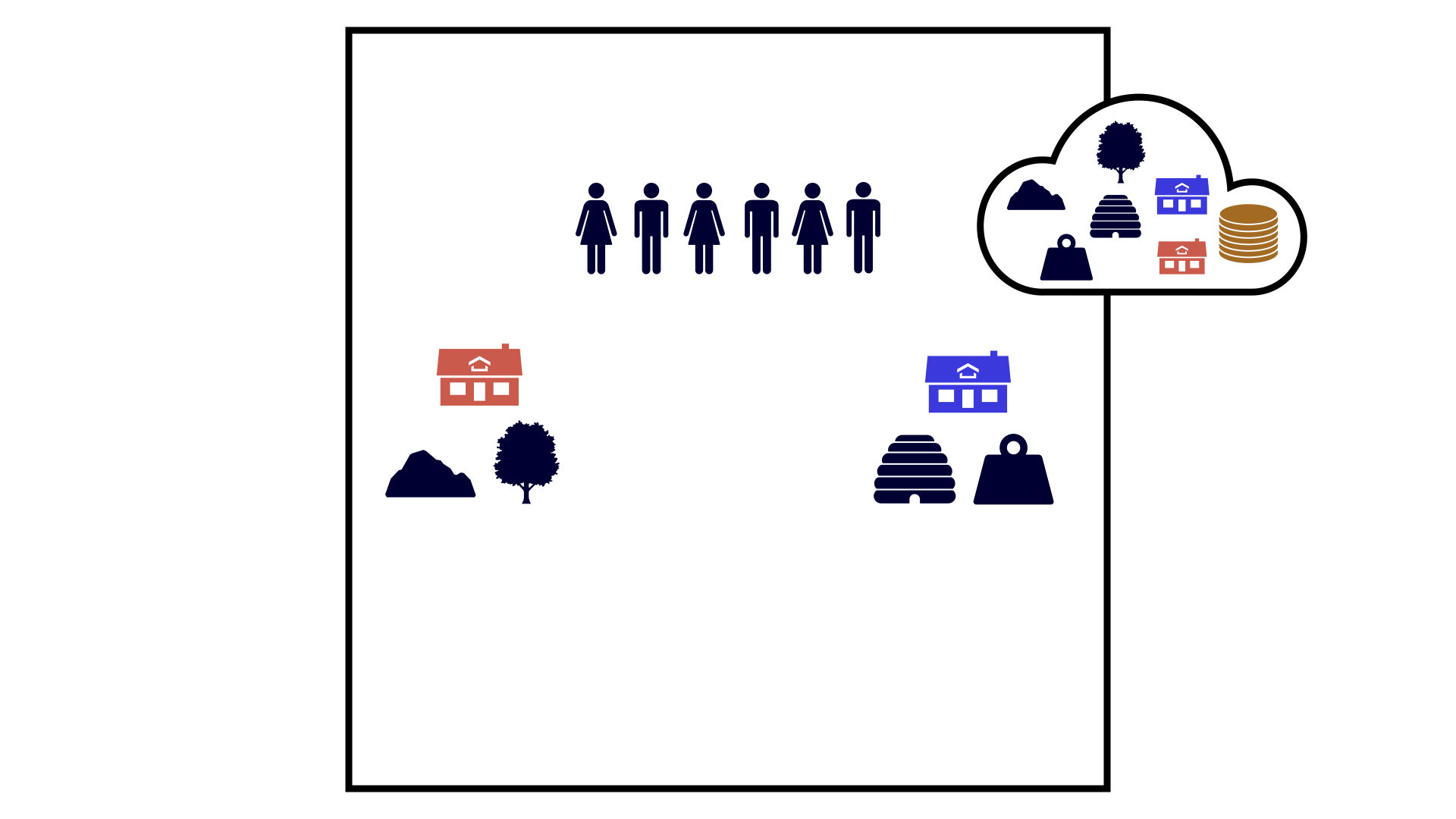}
	\caption{A schematic figure showing the environment of the Modified AI-Economist with Communication/Teaching used in this paper. In all simulations of this paper, there are 6 agents in the environment which simultaneously cooperate and compete to gather and trade four natural resources, using them to build houses alone or together \textendash via communication or teaching \textendash and earn incomes, and at the end of each tax period, pay their taxes to the central planner. The central planner optimizes its own reward function which could be a combination of equality and productivity in the society, and returns an equal division of the total collected taxes to the mobile agents.}
	\label{Figure1}
\end{figure}

Second, the agents are equipped with a voting mechanism and each one of them will have twenty four extra actions to rank the four material resources considering twenty four ranking possibilities. Additionally, all four materials are placed, planted, or extracted randomly in a uniform environment. The agents are initially placed randomly in the environment too. Also, in the modified version of the AI-Economist, the social planner is able to observe the complete public map of the environment (Figs.~\ref{Figure1} and ~\ref{Figure10}). Based on this voting mechanism, across individualistic-collectivistic axis, three different governing systems are introduced. In the Full-Libertarian system, the social planner determines the tax rates considering a particular social welfare function \textendash such as inverse income weighted utility or the multiplication of equality and productivity \textendash and the policy network of the agents now produces an action ranking four different resources. Then, the agents can invest individually their taxes on planting or extraction rates of each one of the four material resources considering how they rank them. In the Semi-Libertarian/Utilitarian system, the tax rates are optimized by the social planner again considering a particular social welfare function. Moreover, the policy network of the agents, as before, produces an action ranking the four material resources. However, the social planner in this case uses the Borda vote counting method to rank the four material types based on the votes of all agents. Then the social planner invests the collected taxes on planting or extraction of the four resources based on the counted votes of all agents. Finally, in the Full-Utilitarian system, the social planner simultaneously optimizes the tax rates and the ranking order of all four materials considering again a suitable social welfare function. Then, it invests the collected taxes accordingly on the planting or extraction rates of all four materials.

Third, communication and teaching capabilities are added to the environment through two variants of signaling game with possible misalignment. When agents decide to build each one of two house types together, they need to communicate the kind of required materials that they have to build that specific house type via four alphabetic letters: [a, b, c, d]. [a, b] refer to the materials required for building a red house type such as wood and stone, while [c, d] refer to the materials required for building a blue house type such as iron and soil. In the Modified AI-Economist with Communication, they are six agents and their language is maximally misaligned in the beginning of each simulation ([a, b, c, d], [d, a, b, c], [c, d, a, b], [b, c, d, a], [d, c, b, a], and [b, a, d, c]). Here, if the agents both use the same kind of letters for both required material resources, they are able to build a house and obtain a large reward. Otherwise, they correct one misaligned letter and obtain a small reward. This process continues until the language of all agents are aligned together, or an episode is ended. In the Modified AI-Economist with Teaching, three agents are teachers and three agents are students. The language of teachers is fixed across training to [a, b, c, d], while again initially the language of students are set to be maximally misaligned ([d, a, b, c], [c, d, a, b], and [b, c, d, a]). Here, a teacher decides to teach building a house type to a student so they both would be able to build a house and share a large reward. In this case, the language of teachers is a reference language which the language of students is compared against. If a pair of teacher and student use the same set of two letters for the required two material resources, they would be able to build that kind of house type. Otherwise, one letter of the language of the student is modified to match the language of the teacher and they obtain a small reward. This process continues until the language of all students are matched to the language of the teachers, or an episode is ended. Finally, the language alignment of agents are calculated across an episode as the average letter aligment of two sequence of language letters among all possible pairs of agents.

\begin{figure}
	\centering
	\includegraphics[width=0.7\linewidth]{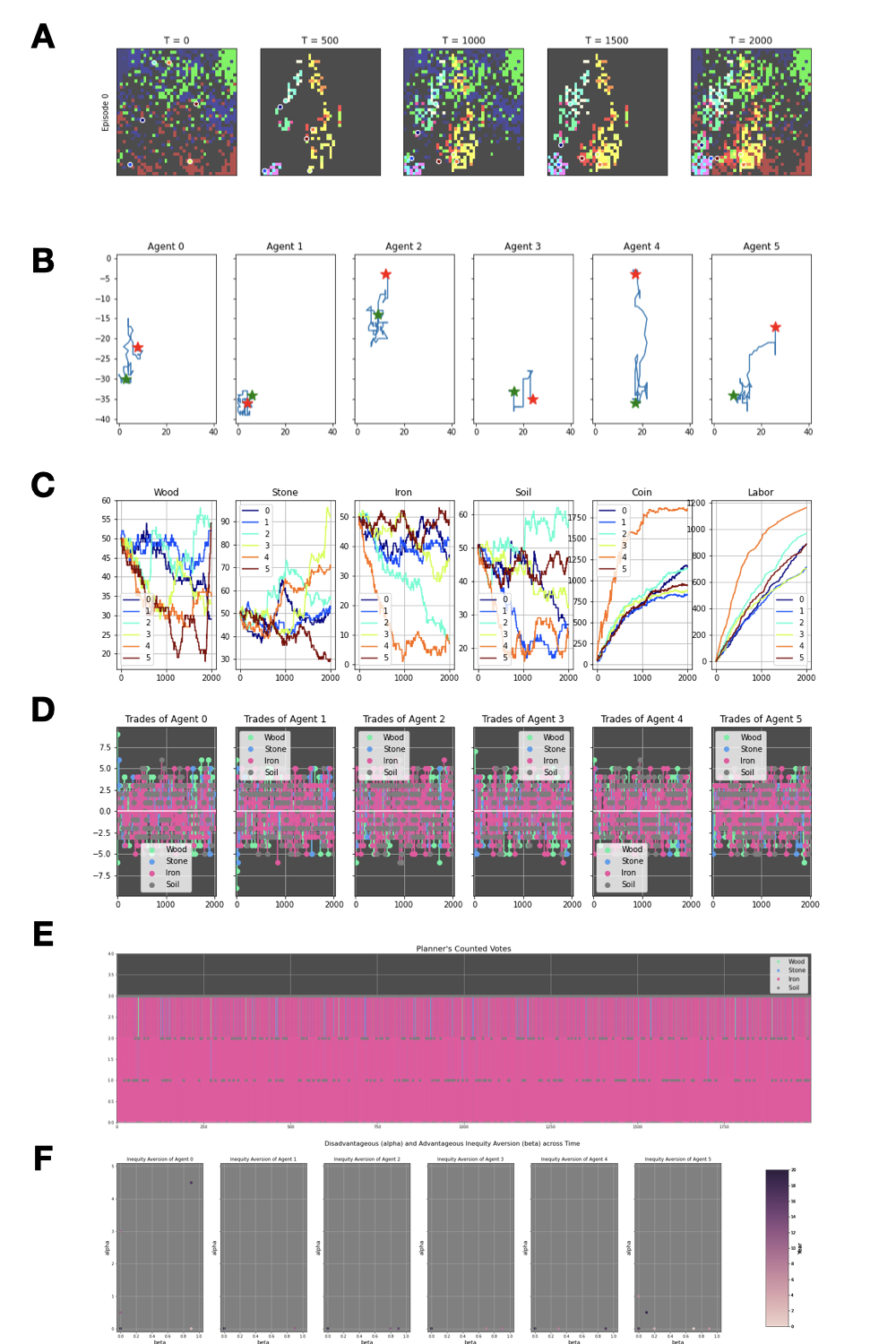}
	\caption{Sample plots obtained from running the Modified AI-Economist with Communication under Semi-Libertarian/Utilitarian governing system with equality times productivity as the objective function of the central planner. (A) The environment across five time-points of an episode, (B) the movement of the agents across an episode, (C) the budgets of four resources plus coin and labor of the agents across an episode, (D) the trades of four resources of the agents across an episode, (E) the counted votes of the agents across an episode, (F) and the inequity aversion coefficients of the agents across an episode.}
	\label{Figure2}
\end{figure}

\begin{figure}
	\centering
	\includegraphics[width=0.7\linewidth]{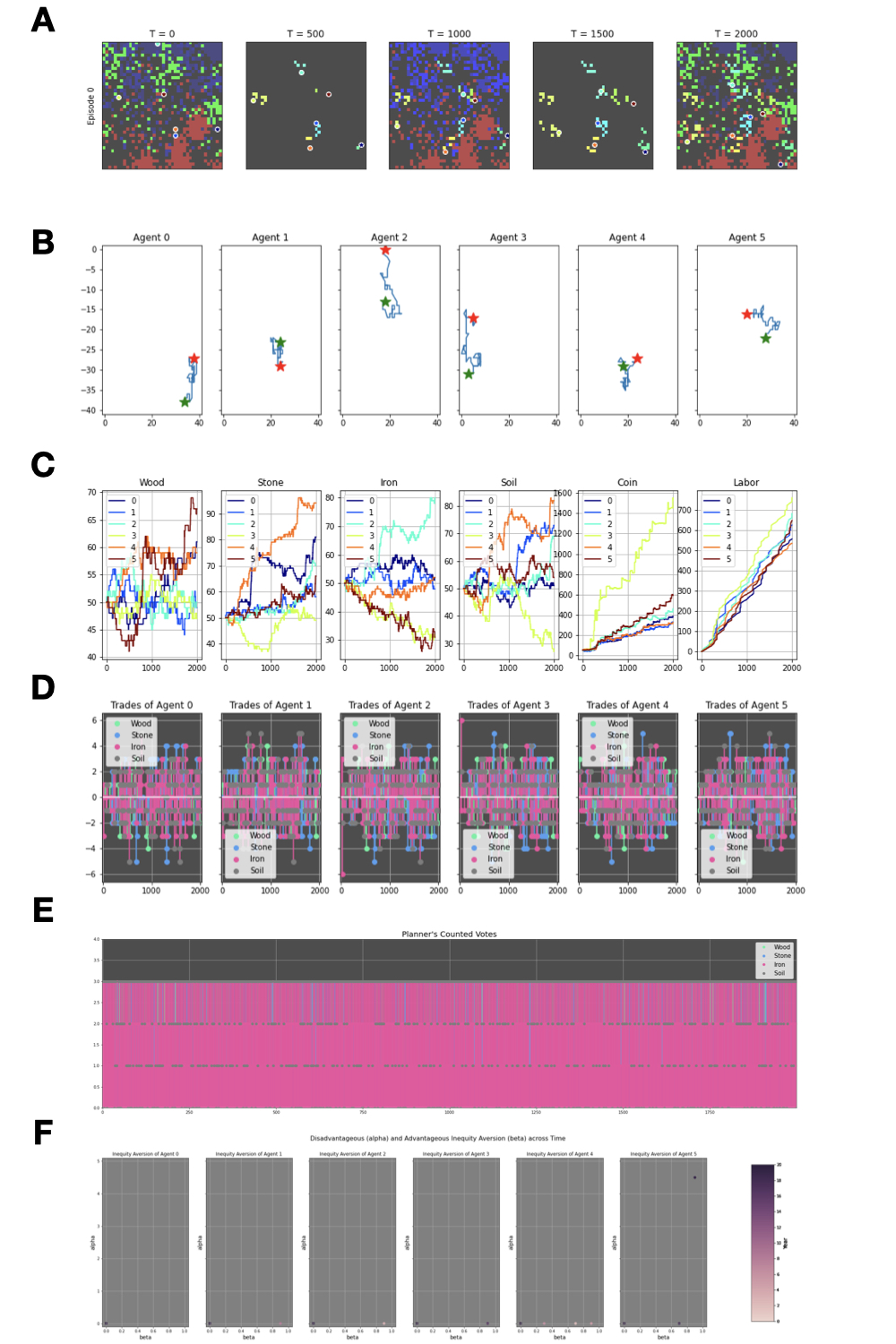}
	\caption{Sample plots obtained from running the Modified AI-Economist with Teaching under Semi-Libertarian/Utilitarian governing system with equality times productivity as the objective function of the central planner. (A) The environment across five time-points of an episode, (B) the movement of the agents across an episode, (C) the budgets of four resources plus coin and labor of the agents across an episode, (D) the trades of four resources of the agents across an episode, (E) the counted votes of the agents across an episode, (F) and the inequity aversion coefficients of the agents across an episode.}
	\label{Figure3}
\end{figure}

Moreover, as a post-processing step, a simple model of morality is developed to compute the effects of language alignment under various governing systems on the morality of the agents in the environment.

Morality can be defined as a suit of cognitive mechanisms makes an agent less selfish, thus to divide the benefits of a cooperation fairly \citep{Greene2013}. In this paper, considering the current framework of the AI-Economist, fairness has been modeled as a self-centered inequity aversion which is a simple but well-known model in economics \citep{Camerer2003, Fehr1999, Hughes2018}. Self-centered inequity aversion means that people resist inequitable outcomes \textendash comparing themselves to others \textendash and they are willing to give up some material payoff to move in the direction of more equitable outcomes. It has been shown that in the presence of some inequity averse people, \textit{fair} and \textit{cooperative} as well as \textit{competitive} and \textit{non-cooperative} behavioral patterns can be explained in a coherent framework. An interesting feature of this examination is that the heterogeneity of preferences interacts in important ways with the economic environment. It has been shown, in particular, that the economic environment determines the preference type that is decisive for the prevailing behavior in equilibrium. This means that, in the presence of heterogeneous preferences, the economic environment has a whole new dimension of effects \citep{Fehr1999}. Thus this method is a natural choice to reveal the impacts of governing systems on the emergence of fairness in the society.

The original inequity averse utility function is as follows, in which \( r_{1}, ..., r_{N} \) is the extrinsic payoffs achieved by each of \( N \) agents, and the additional terms are intrinsic payoffs due to emergence of inequity averse features in the agents \citep{Fehr1999}:
\begin{equation}\label{Equation1}
	U_i(r_{i}, ..., r_{N}) = r_{i} - \frac{\alpha_{i}}{N - 1} \sum_{j \neq i} \max(r_{j} - r_{i}, 0) - \frac{\beta_{i}}{N - 1} \sum_{j \neq i}  \max (r_{i} - r_{j}, 0)
\end{equation}
The parameter \( \alpha_{i} \) controls an agent’s aversion to disadvantageous inequity and could be greater than one. A larger value for \( \alpha_{i} \) implies a larger utility loss when other agents achieve rewards greater than one’s own. Likewise, the parameter \( \beta_{i} \) controls an agent’s aversion to advantageous inequity, utility lost when performing better than others and it is assumed to be less than one. Though here is some empirical support for \( \alpha_{i} \geq \beta_{i} \) \textendash most people are loss averse in social comparisons \textendash the evidence is mixed \citep{Hughes2018}. One aim of this project is to show how to calculate the values of these parameters for different governing systems.

To adopt the above equation for Markov games where the rewards of different players may occur on different time-steps, the key step is to introduce per-player temporal smoothing of the reward traces \citep{Hughes2018}. In this case, let \( r_{i}(s, a) \) denotes the reward obtained by the \( i \)-th player when it takes action \( a \) from state \( s \). For convenience, it could be written with a time index: \( r_{i}^{t} := r_{i} (s^{t}, a^{t}) \). The subjective reward in the original formulation of inequity aversion for Markov games \citep{Hughes2018} is denoted as \( u_{i}(s, a) \), received by the \( i \)-th player when it takes action \( a \) from state \( s \) to be:
\begin{equation}\label{Equation2}
	u_{i} (s_{i}^{t}, a_{i}^{t}) = r_{i} (s_{i}^{t}, a_{i}^{t}) - \frac{\alpha_{i}}{N - 1} \sum_{j \neq i} \max(e_{j}^{t} (s_{j}^{t}, a_{j}^{t}) - e_{i}^{t} (s_{i}^{t}, a_{i}^{t}), 0) - \frac{\beta_{i}}{N - 1} \sum_{j \neq i} \max(e_{i}^{t} (s_{i}^{t}, a_{i}^{t}) - e_{j}^{t} (s_{j}^{t}, a_{j}^{t}), 0)
\end{equation}
where the temporal smoothed rewards \( e_{j}^{t} \) for the agents \( j = 1, ..., N \) are updated at each time-step \( t \) according to:
\begin{equation}\label{Equation3}
	e_{j}^{t} (s_{j}^{t}, a_{j}^{t}) =  \gamma \lambda e_{j}^{t - 1} (s_{j}^{t - 1}, a_{j}^{t - 1}) + r_{j}^{t} (s_{j}^{t}, a_{j}^{t})
\end{equation}
where \( \gamma \) is the discount factor and \( \lambda \) is a hyper-parameter similar to mathematical formulation of eligibility traces \citep{Sutton2018}.

In Eq.~\ref{Equation2}, normally the reward function of the reinforcement learning agent is set to \( u_{i} (s_{i}^{t}, a_{i}^{t}) \), and some values for parameters \textendash \( \alpha_{i} \), \( \beta_{i} \), and \( \lambda \) \textendash are assumed in advance. However, here, it is not assumed beforehand the existence of prosocial or non-prosocial agents with known parameters. Instead, the emergence of fair \textendash moral or cooperative \textendash behaviors under different governing systems is examined by calculating these parameters from available training data. To this end, Eq.~\ref{Equation2} is reformulated as follows:
\begin{equation}\label{Equation4}
	r_{i} (s_{i}^{t}, a_{i}^{t}) = e_{i} (s_{i}^{t}, a_{i}^{t}) - \frac{\alpha_{i}}{N - 1} \sum_{j \neq i} \max(e_{j}^{t} (s_{j}^{t}, a_{j}^{t}) - e_{i}^{t} (s_{i}^{t}, a_{i}^{t}), 0) - \frac{\beta_{i}}{N - 1} \sum_{j \neq i} \max(e_{i}^{t} (s_{i}^{t}, a_{i}^{t}) - e_{j}^{t} (s_{j}^{t}, a_{j}^{t}), 0)
\end{equation}

In Eq.~\ref{Equation4}, on the left side, the \( r_{i} (s_{i}^{t}, a_{i}^{t}) \) refers to many instances of the reward function of a single agent optimized during the MARL. Also, on the right side, as before, the \( e_{i} (s_{i}^{t}, a_{i}^{t}) \) refers to the temporally smoothed reward function having one free parameter \( \lambda \) which could be set beforehand and here I use \( \lambda = 0.5 \). Basically, by fitting many available training data-points of MARL for a single agent \textendash using Eq.~\ref{Equation4} \textendash it is possible to find the best values for \( \alpha_{i} \) and \( \beta_{i} \). In this paper, it is assumed that \( 0 < \alpha_{i} < 5 \) and \( 0 < \beta_{i} < 1 \) (i.e. agents are more disadvantageous inequity averse than advantageous inequity averse) and a grid search is used by dividing these intervals uniformly (for \( \alpha_{i} \) in intervals of 0.5 and for \( \beta_{i} \) in intervals of 0.1) to find optimized values for \( \alpha_{i} \) and \( \beta_{i} \). These parameters might change across an episode or during training \textendash i.e. the agents could be getting more or less fair or moral as an episode or training continues. The whole point here is that the training environment due to the governing systems will determine how these parameters change, and how the moral behaviors of agents co-evolve with the governing system of their environment.

\section{Results}
\begin{figure}
	\centering
	\includegraphics[width=0.7\linewidth]{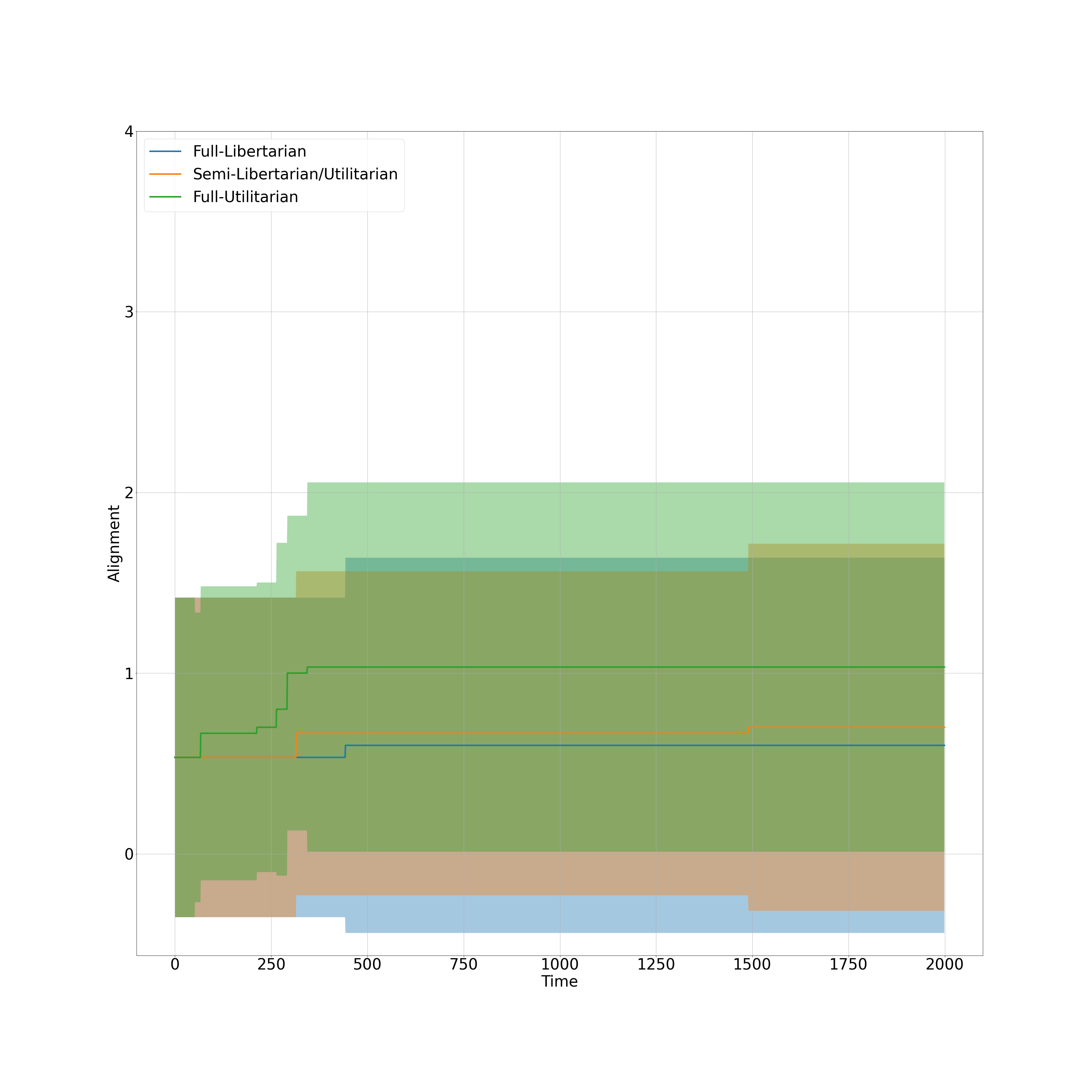}
	\caption{The language alignment across an episode for three governing systems of the Modified AI-Economist with Communication. As it is clear from this plot, a collectivistic government such as Full-Utilitarian governing system is more favourable for the evolution of language alignment compared to an individualistic government such as Full-Libertarian governing system. However, full alignment \textendash which is equal to 4 \textendash does not happen under any of these governing systems, thus none of the agents can build houses together.}
	\label{Figure4}
\end{figure}

\begin{figure}
	\centering
	\includegraphics[width=0.7\linewidth]{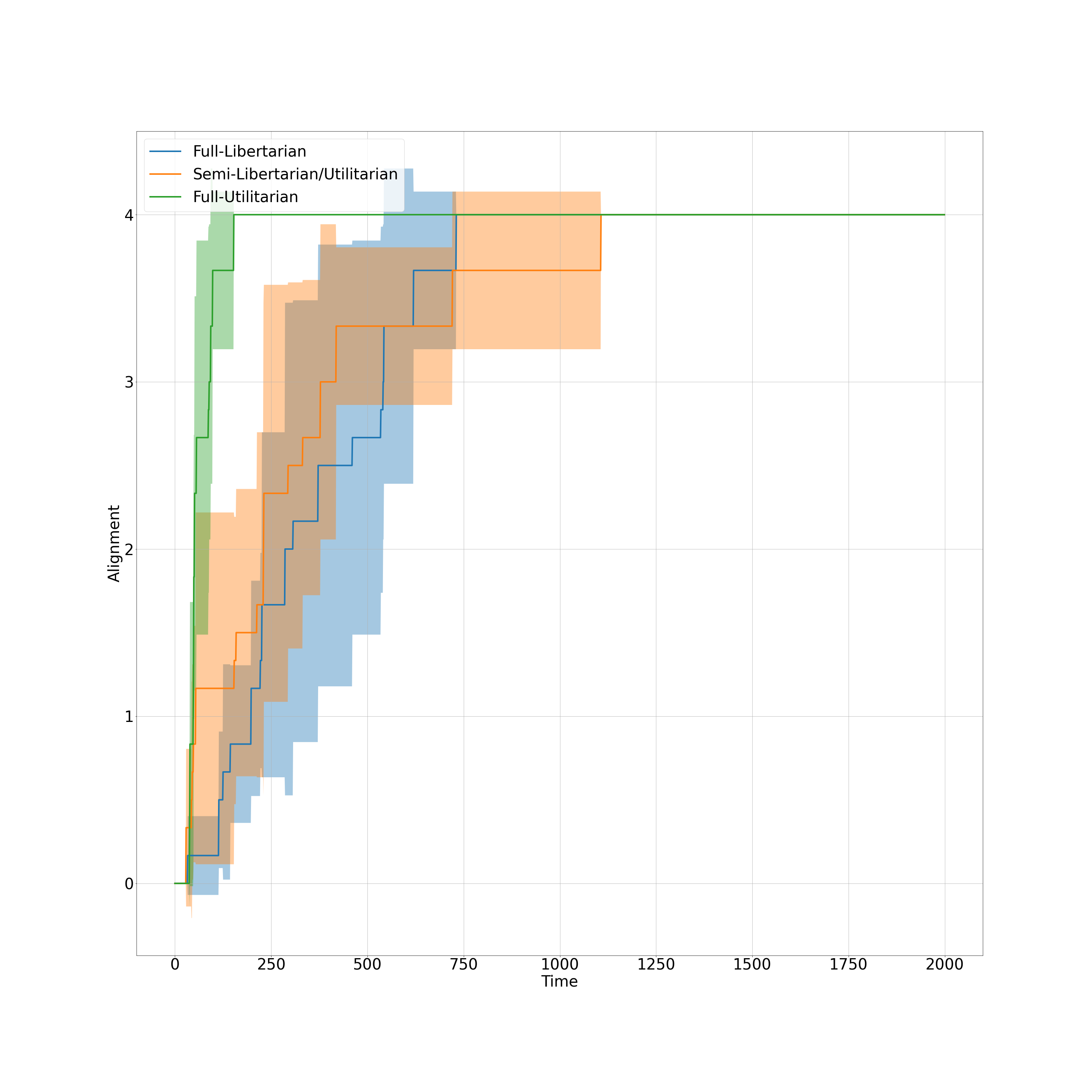}
	\caption{The language alignment across an episode for three governing systems of the Modified AI-Economist with Teaching. As it is clear from this plot, a collectivistic government such as Full-Utilitarian governing system is more favourable for the evolution of language alignment compared to an individualistic government such as Full-Libertarian governing system. In this case, full alignment \textendash which is equal to 4 \textendash happens under all three governing systems, thus the agents can build houses together under all these governing systems.}
	\label{Figure5}
\end{figure}

Fig.~\ref{Figure4} shows the language alignment across three governing systems \textendash Full-Libertarian, Semi-Libertarian/Utilitarian, and Full-Utilitarian \textendash for the Modified AI-Economist with Communication. This figure shows that the rate of language alignment under a collectivistic governing system such as Full-Utilitarian government is higher that the rate of language alignment under an individualistic governing system such as Full-Libertarian government. However, full-alignment never happens under all three governing systems. Moreover, Fig.~\ref{Figure5} shows the language alignment across three governing systems \textendash Full-Libertarian, Semi-Libertarian/Utilitarian, and Full-Utilitarian \textendash for the Modified AI-Economist with Teaching. This figure again shows that the rate of language alignment under a collectivistic governing system such as Full-Utilitarian government is higher that the rate of language alignment under an individualistic governing system such as Full-Libertarian government. In this case, full-alignment happens under all three governing systems, but happens faster under more collectivistic government. 

\begin{figure}
	\centering
	\includegraphics[width=0.7\linewidth]{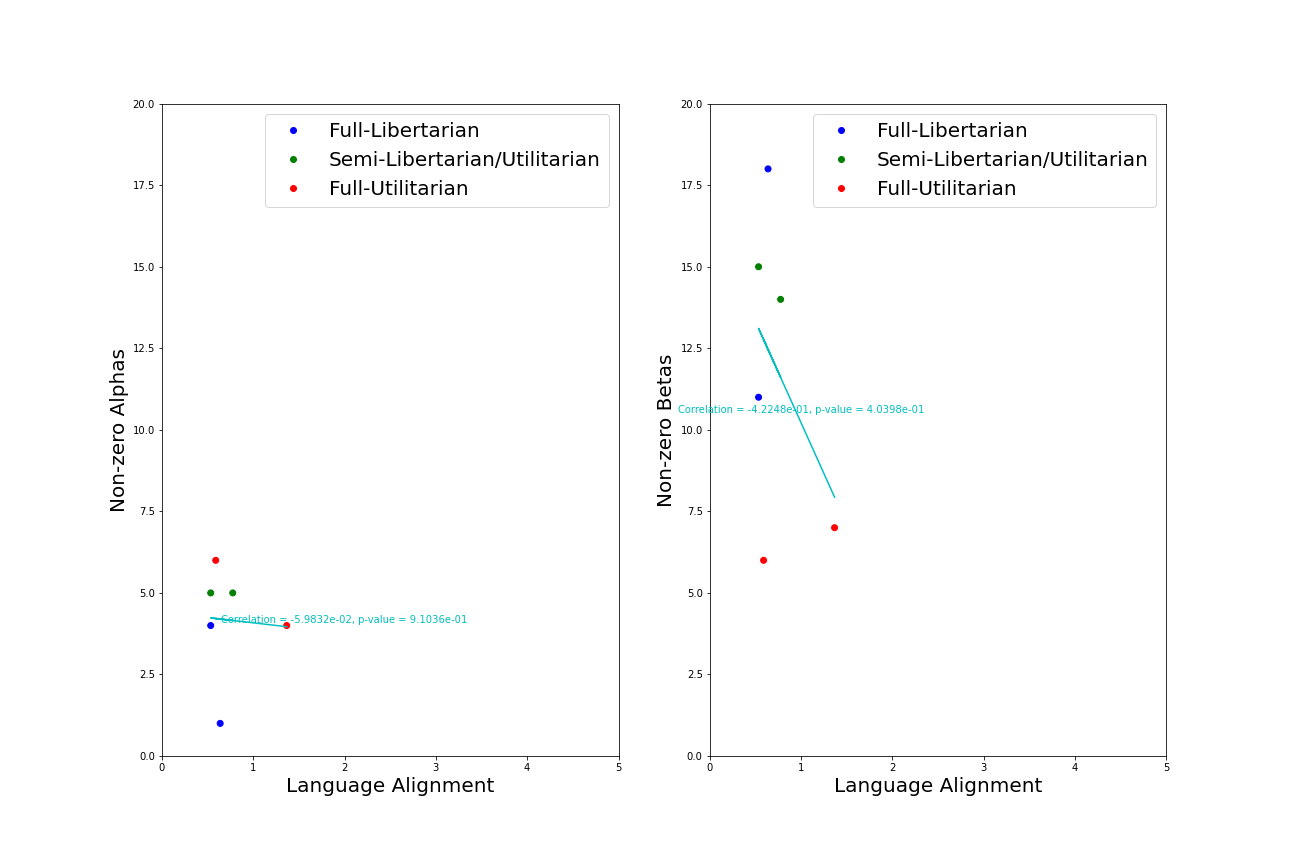}
	\caption{Inequity aversion coefficients versus language alignment across three governing systems for the Modified AI-Economist with Communication. Here, full-alignment does not happen for any pair of agents under all three governing systems. As it is clear, from these plots, almost there is not any correlation between language alignment and alpha \textendash the disadvantageous inequity aversion coefficient. On the other hand, beta \textendash the advantageous inequity aversion coefficient  \textendash has a negative correlation with language alignment.}
	\label{Figure6}
\end{figure}

\begin{figure}
	\centering
	\includegraphics[width=0.7\linewidth]{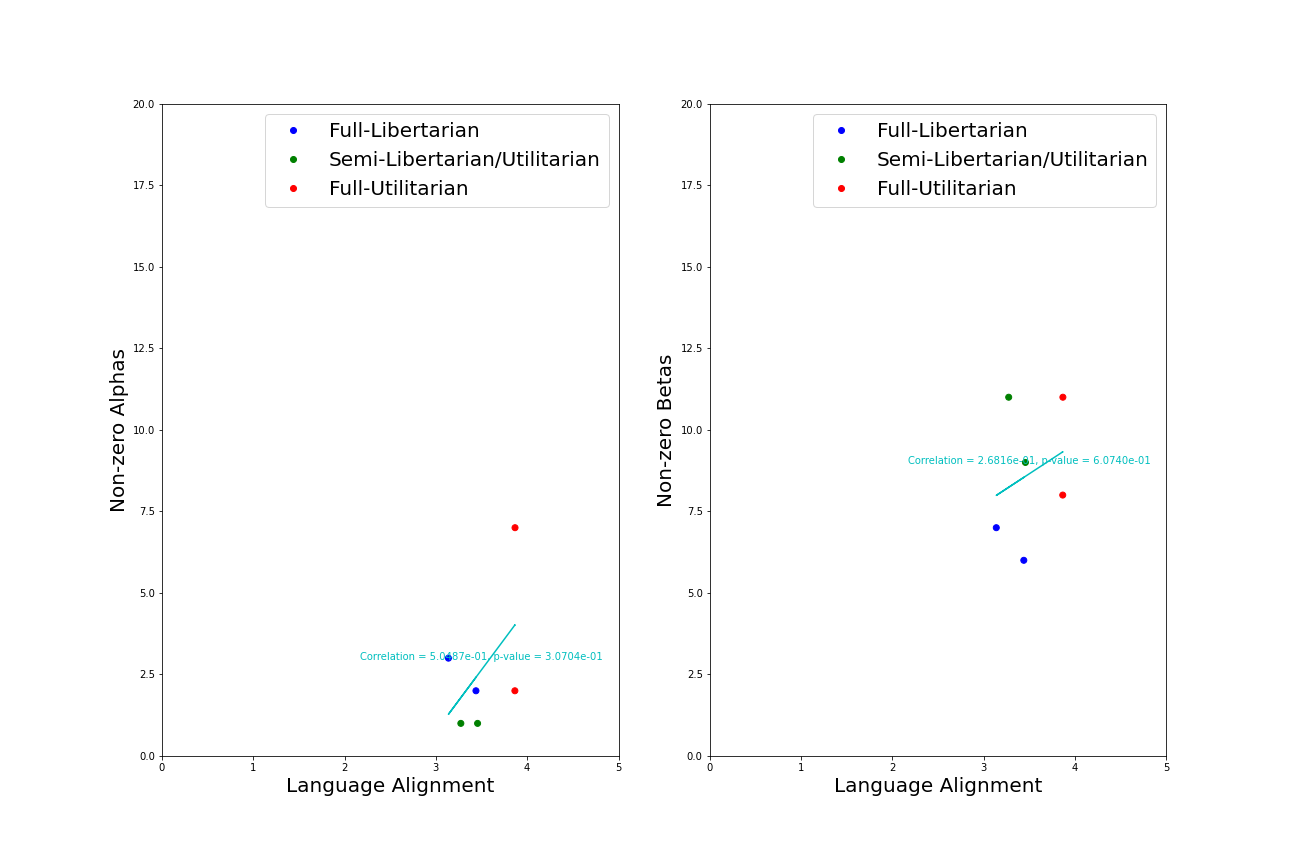}
	\caption{Inequity aversion coeffcients versus language alignment across three governing systems for the Modified AI-Economist with Teaching. Here, full-alignment happens under all three governing systems. As it is clear, from these plots, both alpha \textendash the disadvantageous inequity aversion coefficient \textendash and beta \textendash the advantageous inequity aversion coefficient  \textendash has positive correlations with language alignment.}
	\label{Figure7}
\end{figure}

Furthermore, Fig.~\ref{Figure6} shows the inequity aversion coefficients \textendash alpha and beta \textendash versus language alignment across three governing systems in the Modified AI-Economist with Communication. As it clear from these plots, there is almost no correlation between the disadvantageous inequity aversion \textendash alpha \textendash and language alignment, while the advantageous inequity aversion \textendash beta \textendash has a negative correlation with language alignment. We should interprete this result in the light of the fact that, in this case, full-alignment between any pair of agents never happens. On the other hand, Fig.~\ref{Figure7} shows the inequity aversion coefficients \textendash alpha and beta \textendash versus language alignment across three governing systems in the Modified AI-Economist with Teaching. As it is clear from these plots, here both disadvantageous \textendash alpha \textendash and advantageous \textendash beta \textendash inequity aversion coefficients have positive correlations with language alignment. In this case, the full-alignment of language also happens soon in the episode under all three governing systems. I believe this and previous results provide some evidence for, at least, positive correlation of the advantageous inequity aversion coefficient with evolution of language alignment.

\begin{figure}
	\centering
	\includegraphics[width=0.7\linewidth]{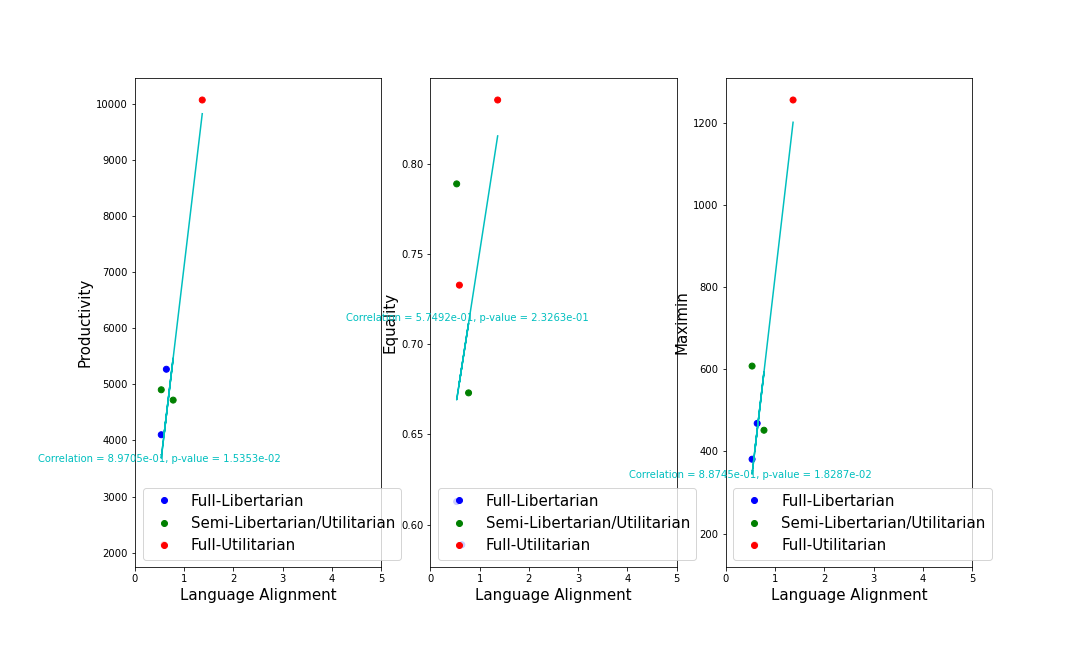}
	\caption{Productivity, Equality, and Maximin versus language alignment across three governing systems for the Modified AI-Economist with Communication. As it is clear from these plots, all three quantities have positive correlations with language alignment.}
	\label{Figure8}
\end{figure}

\begin{figure}
	\centering
	\includegraphics[width=0.7\linewidth]{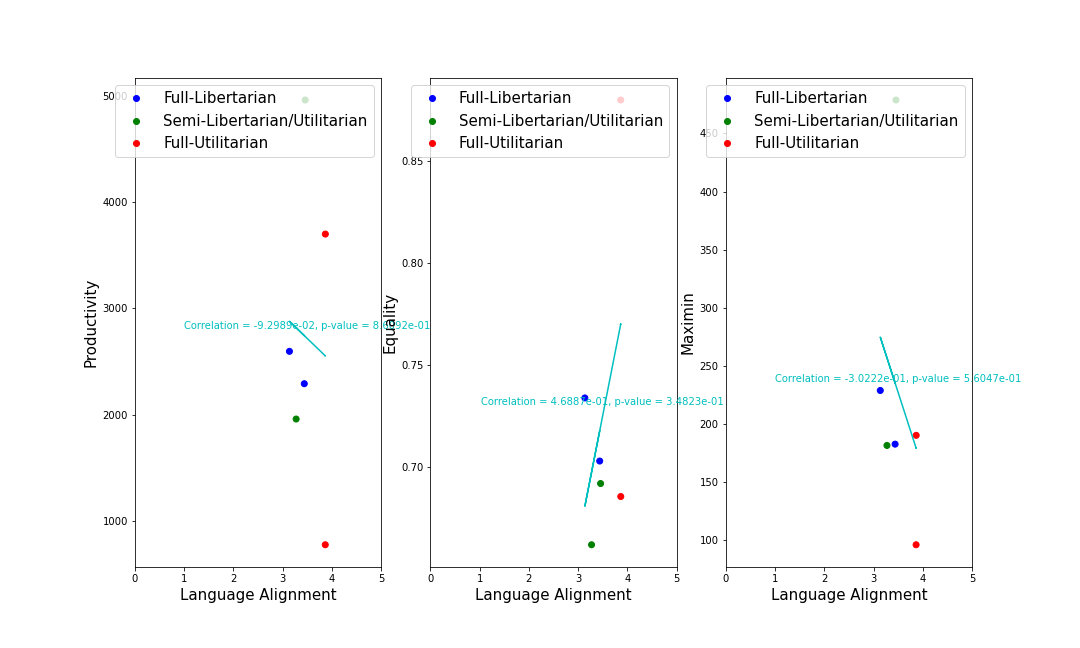}
	\caption{Productivity, Equality, and Maximin versus language alignment across three governing systems for the Modified AI-Economist with Teaching. As it is clear from these plots, Productivity and Maximin have negative correlations, while Equality has a positive correlation with language alignment.}
	\label{Figure9}
\end{figure}

\begin{figure}
	\centering
	\includegraphics[width=0.7\linewidth]{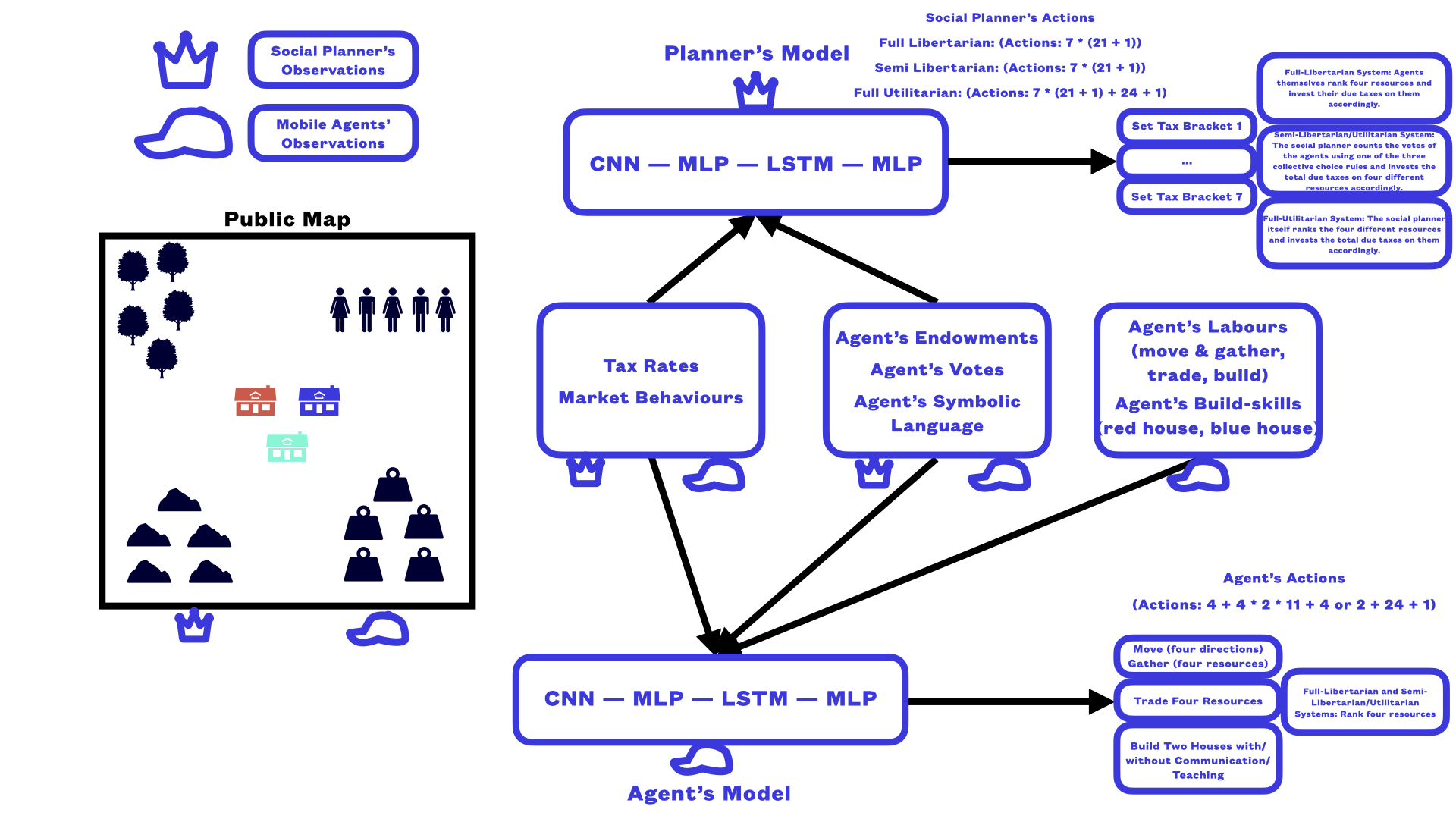}
	\caption{Observation and action spaces for economic agents and the social planner. The agents and the planner observe different subsets of the world state. Agents observe their spatial neighborhood, market prices, tax rates, inventories, votes, symbolic language, labor, and skill level. Agents can decide to move (and therefore gather if moving onto a resource), buy, sell, build, vote, or communicate. There are maximum 121 (communication) and 119 (teaching) unique actions available to the agents. The planner observes the public spatial map, market prices, tax rates, agent inventories, votes, and symbolic language. The social planner in both environments decides how to set tax rates, choosing one of 22 settings for each of the 7 tax brackets. MLP: multi-layer perceptron, LSTM: long short-term memory, CNN: convolutional neural network. This figure should be compared to Fig. 9 of the original AI-Economist paper (\cite{Zheng2022}).}
	\label{Figure10}
\end{figure}

Finally, Fig.~\ref{Figure8} and Fig.~\ref{Figure9} show Productivity, Equality, and Maximin versus language alignment across three governing systems for the Modified AI-Economist with Communication and with Teaching respectively. In Fig.~\ref{Figure8}, the plots show that there are positive correlations between language alignment and Productivity, Equality, and Maximin. Again, these results should be interpreted cautiously since full-alignment never happens under any of three governing systems. On the other hand, the plots of Fig.~\ref{Figure9} show that there are negative correlations between language alignment and Productivity or Maximin, while there is a positive correlation between language alignment and equality. In this case, this positive correlation can be interpreted as a signature of full-alignment of language which happens under all three governing systems.

While I consider the results of this study tentative, it is worhwhile to ponder about the reasons behind more favourable conditions of Full-Utilitarian government for the evolution of language alignment compared to the Full-Libertarian government in the way that they are modeled in this work. One speculation is that basically among Full-Libertarian, Semi-Libertarian/Utilitarian, and Full-Utilitarian governments as they are modelled in this study, the Full-Utilitarian governing system lacks any individual voting mechanism. If we consider individual voting as a mean of communication and coordination among agents, this lack of individual voting in the Full-Utilitarian government might be the reason behind fast emergence and evolution of language alignment under this type of governing system. Basically, in the current modelling, the agents under Full-Utilitarian society without any means of communication or coordination are motivated to incur high costs and simultaneously gain large rewards by selecting the action to build houses together compared to the societies which have a voting mechanism as a mean of communication and coordination.

\section{Final Remarks}
\paragraph{Current Limitations} There are at least three limitations to the current tentative study. The first limitation comes from the fact that for each set of input parameters of the Modified AI-Economist with Communication or Teaching, only one simulation has been run to generate one set of results. Then a pair of similar runs have been pooled together to have average results across different conditions (Fig.~\ref{Figure12}). Thus, it is reasonable to run multiple times a simulation with a unique set of parameters and then average them all together. The second limitation comes from the fact that communication under the Modified AI-Economist with Communication is a rare event. Basically, the agents have selected the action which needs two agents to build a house very rarely compared to the action which only requires one agent to build a house. This might be resolved by increasing the number of time-steps of the MARL environment, or by increasing the small reward obtained when selecting the rare action for one step of language alignment. Finally, the third limitation is the number of episodes each training has been run which is equal to 5000. As it is mentioned in the introduction of the paper, the optimality of equilibrium selection in MARL is an unsolved problem (\cite{Albrecht2023}). The only way to make sure that the final results are optimal is to let the training of MARL runs for very large number of time-steps. However, here, due to computational resources, the maximum limit of time-steps is chosen to be 5000, which is still a reasonable choice in MARL. (Fig.~\ref{Figure13}) and (Fig.~\ref{Figure14}) show that even with this choice almost all simulations have been converged. Although, the training of the Modified AI-Economist with Teaching is more stable than the training of the Modified AI-Economist with Communication. Considering all these facts, still, it is wise to try more MARL iterations.

\paragraph{Future Directions} Beside above modifications, two other important extensions for this project can be envisioned. The first one is to use more sophisticated modeling of the governing systems, particularly considering the fact that the governing systems in the current format do treat all agents equally. Moreover, they are omniscient and omnipotent meaning that they have complete knowledge of the environment without any noise. Any realistic governmental modelling should include some level of noise in the knowledge and power of the government. The second direction is to use more sophisticated modelling of communication and teaching. In this study, a very simple model of language has been used to model communication and teaching. One important extension of the current study could be more realistic modelling of language in the context of language game (\cite{Karch2023, Ohmer2022}).

\section*{Broader Impact}
As I discussed in my previous papers (\cite{Dizaji2023a, Dizaji2023b}), for the works similar to the current study, it is possible to envision many policy implications, however, we should be cautions about interpreting these results more than what is appropriate, due to many simplifying assumptions inherent in any computational simulation. Overall, the limited modeling framework used in this paper with tentative results shows that in the discussion of the emergence and evolution of language convention, we should cautiously consider the roles of governing systems, and their relations with morality, productivity, equality, and maximin.

For further information, please refer to the Ethics section of the original AI-Economist paper (\cite{Zheng2022}). Particularly, in that section, it has been emphasized on the full-transparency of the codes of a project with a similar scope. As a result, I provided two open Github repositories (\url{https://github.com/aslansd/modified-ai-economist-wc}) (Modified AI-Economist with Communication) and (\url{https://github.com/aslansd/modified-ai-economist-wt}) (Modified AI-Economist with Teaching) having all the required codes, simulations, and notebooks to generate the runs and plots of this paper.

\begin{ack}
\textit{AutocurriculaLab} has been funded in March 2022 and since then has been supported by multiple agencies. Hereby, I acknowledge their supports.
\end{ack}

\medskip
\small

\bibliography{Agent_based_Modelling_for_Complex_Systems_and_Economics}
\bibliographystyle{apalike}

\newpage

\section{Appendix A: The AI-Economist}
Here is a detailed description of the original AI-Economist (\cite{Zheng2022}):

\begin{enumerate}
	\item The AI-Economist is a two-level deep RL framework for policy design in which agents and a social planner co-adapt. In particular, the AI-Economist uses structured curriculum learning to stabilize the challenging two-level, co-adaptive learning problem. This framework has been validated in the domain of taxation. In two-level spatiotemporal economies, the AI-Economist has substantially improved both utilitarian social welfare and the trade-off between equality and productivity over baselines. It was successful to do this despite emergent tax-gaming strategies, accounting the emergent labor specialization, agent interactions, and behavioral changes.
	
	\item Stabilizing the training process in two-level RL is difficult. To overcome, the training procedure in the AI-Economist has two important features - curriculum learning and entropy-based regularization. Both of them encourage the agents and the social planner to co-adopt gradually and not stopping exploration too early during training and getting trapped in local minima. Furthermore, the AI-Economist is a game of imperfect (the agents and the social planner do not have access to the perfect state of the world) but complete (the agents and the social planner know the exact rules of the game) information.
	
	\item The Gather-Trade-Build economy of the AI-Economist is a two-dimensional spatiotemporal economy with agents who move, gather resources (stone and wood), trade them, and build houses. Each agent has a varied house build-skill which sets how much income an agent receives from building a house. Build-skill is distributed according to a Pareto distribution. As a result, the utility-maximizing agents learn to specialize their behaviors based on their build-skill. Agents with low build-skill become gatherers: they earn income by gathering and selling resources. Agents with high build-skill become builders: they learn that it is more profitable to buy resources and then build houses.
	
	\item The Open-Quadrant environment of the Gather-Trade-Build economy has four regions delineated by impassable water with passageways connecting each quadrant. Quadrants contain different combinations of resources: both stone and wood, only stone, only wood, or nothing. Agents can freely access all quadrants, if not blocked by objects or other agents. This scenario uses a fixed set of build-skill based on a clipped Pareto distribution and determine each agent’s starting location based on its assigned build-skill. The Open-Quadrant scenario assigns agents to a particular corner of the map, with similarly skilled agents being placed in the same starting quadrant (Agents in the lowest build-skill quartile start in the wood quadrant; those in the second quartile start in the stone quadrant; those in the third quartile start in the quadrant with both resources; and agents in the highest build-skill quartile start in the empty quadrant).
	
	\item The state of the world is represented as an \( n_{h} \times n_{w} \times n_{c} \) tensor, where \( n_{h} \) and \( n_{w} \) are the size of the world and \( n_{c} \) is the number of unique entities that may occupy a cell, and the value of a given element indicates which entity is occupying the associated location. The action space of the agents includes four movement actions: up, down, left, and right. Agents are restricted from moving onto cells that are occupied by another agent, a water tile, or another agent’s house. Stone and wood stochastically spawn on special resource regeneration cells. Agents can gather these resources by moving to populated resource cells. After harvesting, resource cells remain empty until new resources spawn. By default, agents collect one resource unit, with the possibility of a bonus unit also being collected, the probability of which is determined by the agent’s gather-skill. Resources and coins are accounted for in each agent’s endowment \( x \), which represents how many coins, stone, and wood each agent owns.
	
	\item Agent's observations include the state of their own endowment (wood, stone, and coin), their own build-skill level, and a view of the world state tensor within an egocentric spatial window. The experiment use a world of 25 by 25 for 4-agent and 40 by 40 for 10-agent environments, where agent spatial observations have size 11 by 11 and are padded as needed when the observation window extends beyond the world grid. The planner observations include each agent’s endowment but not build-skill level. The planner does not observe the spatial state of the world.
	
	\item Agents can buy and sell resources from one another through a continuous double-auction. Agents can submit asks (the number of coins they are willing to accept) or bids (how much they are willing to pay) in exchange for one unit of wood or stone. The action space of the agents includes 44 actions for trading, representing the combination of 11 price levels (0, ..., 10 coins), 2 directions (bids and asks), and 2 resources (wood and stone). Each trade action maps to a single order (i.e., bid three coins for one wood, ask for five coins in exchange for one stone, etc.). Once an order is submitted, it remains open until either it is matched (in which case a trade occurs) or it expires (after 50 time steps). Agents are restricted from having more than five open orders for each resource and are restricted from placing orders that they cannot complete (they cannot bid with more coins than they have and cannot submit asks for resources that they do not have). A bid/ask pair forms a valid trade if they are for the same resource and the bid price matches or exceeds the ask price. When a new order is received, it is compared against complementary orders to identify potential valid trades. When a single bid (ask) could be paired with multiple existing asks (bids), priority is given to the ask (bid) with the lowest (highest) price; in the event of ties, priority then is given to the earliest order and then at random. Once a match is identified, the trade is executed using the price of whichever order was placed first. For example, if the market receives a new bid that offers eight coins for one stone and the market has two open asks offering one stone for three coins and one stone for seven coins, received in that order, the market would pair the bid with the first ask and a trade would be executed for one stone at a price of three coins. The bidder loses three coins and gains one stone; the asker loses one stone and gains three coins. Once a bid and ask are paired and the trade is executed, both orders are removed. The state of the market is captured by the number of outstanding bids and asks at each price level for each resource. Agents observe these counts for both their own bids/asks and the cumulative bids/asks of other agents. The planner observes the cumulative bids/asks of all agents. In addition, both the agents and the planner observe historical information from the market: the average trading price for each resource, as well as the number of trades at each price level.
	
	\item Agents can choose to spend one unit of wood and one unit of stone to build a house, and this places a house tile at the agent’s current location and earns the agent some number of coins. Agents are restricted from building on source cells as well as locations where a house already exists. The number of coins earned per house is identical to an agent’s build-skill, a numeric value between 10 and 30. Hence, agents can earn between 10 and 30 coins per house built. Build-skill is heterogeneous across agents and does not change during an episode. Each agent’s action space includes one action for building. Over the course of an episode of 1000 time steps, agents accumulate labor cost, which reflects the amount of effort associated with their actions. Each type of action (moving, gathering, trading, and building) is associated with a specific labor cost. All agents experience the same labor costs.
	
	\item Simulations are run in episodes of 1000 time steps, which is subdivided into 10 tax periods or tax years, each lasting 100 time steps. Taxation is implemented using income brackets and bracket tax rates. All taxation is anonymous: Tax rates and brackets do not depend on the identity of taxpayers. On the first time step of each tax year, the marginal tax rates are set that will be used to collect taxes when the tax year ends. For taxes controlled by the deep neural network of the social planner, the action space of the planner is divided into 7 action subspaces, one for each tax bracket: \( (0, 0.05, 0.10, ..., 1.0)^{7} \). Each subspace denotes the set of discretized marginal tax rates available to the planner. Discretization of tax rates only applies to deep learning networks, enabling standard techniques for RL with discrete actions. The income bracket cutoffs are fixed. Each agent observes the current tax rates, indicators of the temporal progress of the current tax year, and the set of sorted and anonymized incomes the agents reported in the previous tax year. In addition to this global tax information, each agent also observes the marginal rate at the level of income it has earned within the current tax year so far. The planner also observes this global tax information, as well as the non-anonymized incomes and marginal tax rate (at these incomes) of each agent in the previous tax year.
	
	\item The payable tax for income \( z \) is computed as follows:
	\begin{equation}\label{Equation5}
		T(z) = \sum_{j = 1}^{B} \tau_{j} \cdot ((b_{j + 1} - b_{j}) \mathbf{1} [z > b_{j + 1}] + (z - b_{j}) \mathbf{1} [b_{j} < z \leq b_{j + 1}])
	\end{equation}
	where \( B \) is the number of brackets, and \( \tau_{j} \) and \( b_{j} \) are marginal tax rates and income boundaries of the brackets, respectively.
	
	\item An agent’s pretax income \( z_{i} \) for a given tax year is defined simply as the change in its coin endowment \( C_{i} \) since the start of the year. Accordingly, taxes are collected at the end of each tax year by subtracting \( T(z_{i}) \) from \( C_{i} \). Taxes are used to redistribute wealth: the total tax revenue is evenly redistributed back to the agents. In total, at the end of each tax year, the coin endowment for agent \( i \) changes according to \( \bigtriangleup C_{i} = - T(z_{i}) + \frac{1}{N} \sum_{j = 1}^{N} T(z_{j}) \), where \( N \) is the number of agents. Through this mechanism, agents may gain coin when they receive more through redistribution than they pay in taxes. Following optimal taxation theory, agent utilities depend positively on accumulated coin \( C_{i, t} \), which only depends on post-tax income \( \tilde{z}  = z - T(z) \). In contrast, the utility for agent \( i \) depends negatively on accumulated labor \( L_{i, t} = \sum_{k = 0}^{t} l_{i, k} \) at time step \( t \). The utility for an agent \( i \) is:
	\begin{equation}\label{Equation6}
		u_{i, t} = \frac{C^{1 - \eta}_{i, t} - 1}{1 - \eta} - L_{i, t}, \eta > 0
	\end{equation}

	\item Agents learn behaviors that maximize their expected total discounted utility for an episode. It is found that build-skill is a substantial determinant of behavior; agents’ gather-skill empirically does not affect optimal behaviors in our settings. All of the experiments use a fixed set of build-skills, which, along with labor costs, are roughly calibrated so that (i) agents need to be strategic in how they choose to earn income and (ii) the shape of the resulting income distribution roughly matches that of the 2018 U.S. economy with trained optimal agent behaviors.
	
	\item RL provides a flexible way to simultaneously optimize and model the behavioral effects of tax policies. RL is instantiated at two levels, that is, for two types of actors: training agent behavioral policy models and a taxation policy model for the social planner. Each actor’s behavioral policy is trained using deep RL, which learns the weights \( \theta_{i} \) of a neural network \( \pi(a_{i, t} | o_{i, t}; \theta_{i}) \) that maps an actor’s observations to actions. Network weights are trained to maximize the expected total discounted reward of the output actions. Specifically, for an agent \( i \) using a behavioral policy \( \pi_{i}(a_{t} | o_{t}; \theta_{i}) \), the RL training objective is (omitting the tax policy \( \pi_{p} \)):
	\begin{equation}\label{Equation7}
		\max_{\pi_{i}} E_{a_{1} \sim \pi_{1}, ..., a_{N} \sim \pi_{N}, s^{'} \sim P} [\sum^{H}_{t = 0} \gamma^{t} r_{t}]
	\end{equation}
	where \( s^{'} \) is the next state and \( P \) denotes the dynamics of the environment. The objective for the planner policy \( \pi_{p} \) is similar. Standard model-free policy gradient methods update the policy weights \( \theta_{i} \) using (variations of):
	\begin{equation}\label{Equation8}
		\mathbf{\triangle \theta_{i}} \propto E_{{a_{1} \sim \pi_{1}, ..., a_{N} \sim \pi_{N}, s^{'} \sim P}}[\sum^{H}_{t = 0} \gamma^{t} r_{t} \nabla_{\theta_{i}} \log \pi_{i}(a_{i, t} | o_{i, t}; \mathbf{\theta_{i}})]
	\end{equation}

	\item In this work, the proximal policy gradients (PPO) is used to train all actors (both agents and planner). To improve learning efficiency, a single-agent policy network \( \pi(a_{i, t} | o_{i, t}; \theta) \) is trained whose weights are shared by all agents, that is, \( \theta_{i} = \theta \). This network is still able to embed diverse, agent-specific behaviors by conditioning on agent-specific observations.
	
	\item At each time step \( t \), each agent observes the following: its nearby spatial surroundings; its current endowment (stone, wood, and coin); private characteristics, such as its building skill; the state of the markets for trading resources; and a description of the current tax rates. These observations form the inputs to the policy network, which uses a combination of convolutional, fully connected, and recurrent layers to represent spatial, non-spatial, and historical information, respectively. For recurrent components, each agent maintains its own hidden state. The policy network for the social planner follows a similar construction but differs somewhat in the information it observes. Specifically, at each time step, the planner policy observes the following: the current inventories of each agent, the state of the resource markets, and a description of the current tax rates. The planner cannot directly observe private information such as an agent’s skill level.
	
	\item Rational economic agents train their policy \( \pi_{i} \) to optimize their total discounted utility over time while experiencing tax rates \( \tau \) set by the planner’s policy \( \pi_{p} \). The agent training objective is:
	\begin{equation}\label{Equation9}
		\forall i : \max_{\pi_{i}} E_{\tau \sim \pi_{p}, a_{i} \sim \pi_{i}, \mathbf{a_{-i}} \sim \mathbf{\pi_{-i}}, s^{'} \sim P} [\sum^{H}_{t = 1} \gamma^{t} r_{i, t} + u_{i, 0}], r_{i, t} = u_{i, t} - u_{i, t - 1}
	\end{equation}
	where the instantaneous reward \( r_{i, t} \) is the marginal utility for agent \( i \) at time step \( t \). Bold-faced quantities denote vectors, and the subscript \( -i \) denotes quantities for all agents except for \( i \).
	
	\item For an agent population with monetary endowments \( \mathbf{C_{t}} = (C_{1, t}, ..., C_{N, t}) \), the equality \( eq(\mathbf{C_{t}}) \) is defined as:
	\begin{equation}\label{Equation10}
		eq(\mathbf{C}_{t}) = 1 - \frac{N}{N - 1} gini(\mathbf{C}_{t}), 0 \leq eq(\mathbf{C}_{t}) \leq 1
	\end{equation}
	where the Gini index is defined as:
	\begin{equation}\label{Equation11}
		gini(\mathbf{C}_{t}) = \frac{\sum_{i = 1}^{N} \sum_{j = 1}^{N} |C_{i, t} - C_{j, t}|}{2N \sum^{N}_{i = 1} C_{i, t}}, 0 \leq gini(\mathbf{C}_{t}) \leq \frac{N - 1}{N}
	\end{equation}

	\item The productivity is defined as the sum of all incomes:
	\begin{equation}\label{Equation12}
		prod(\mathbf{C}_{t}) = \sum_{i} C_{i, t}
	\end{equation}
	The economy is closed: subsidies are always redistributed evenly among agents, and no tax money leaves the system. Hence, the sum of pretax and post-tax incomes is the same. The planner trains its policy \( \pi_{p} \) to optimize social welfare:
	\begin{equation}\label{Equation13}
		\max_{\pi_{p}} E_{\tau \sim \pi_{p}, \mathbf{a} \sim \mathbf{\pi}, s^{'} \sim P} [\sum^{H}_{t = 1} \gamma^{t} r_{p, t} + swf_{0}], r_{p, t} = swf_{t} - swf_{t - 1}
	\end{equation}

	\item The utilitarian social welfare objective is the family of linear-weighted sums of agent utilities, defined for weights \( \omega_{i} \geq 0 \):
	\begin{equation}\label{Equation14}
		swf_{t} = \sum^{N}_{i = 1} \mathbf{\omega}_{i} \cdot \mathbf{u}_{i, t}
	\end{equation}
	Inverse-income is used as the weights: \( \omega_{i} \propto \frac{1}{C_{i}} \), normalized to sum to one. An objective function is defined that optimizes a trade-off between equality and productivity, defined as the product of equality and productivity:
	\begin{equation}\label{Equation15}
		swf_{t} = eq(\mathbf{C}_{t}) \cdot prod(\mathbf{C}_{t})
	\end{equation}
\end{enumerate}

\newpage

\section{Appendix B: Supplemental Figures}

\newpage

\begin{figure}
	\centering
	\includegraphics[width=0.7\linewidth]{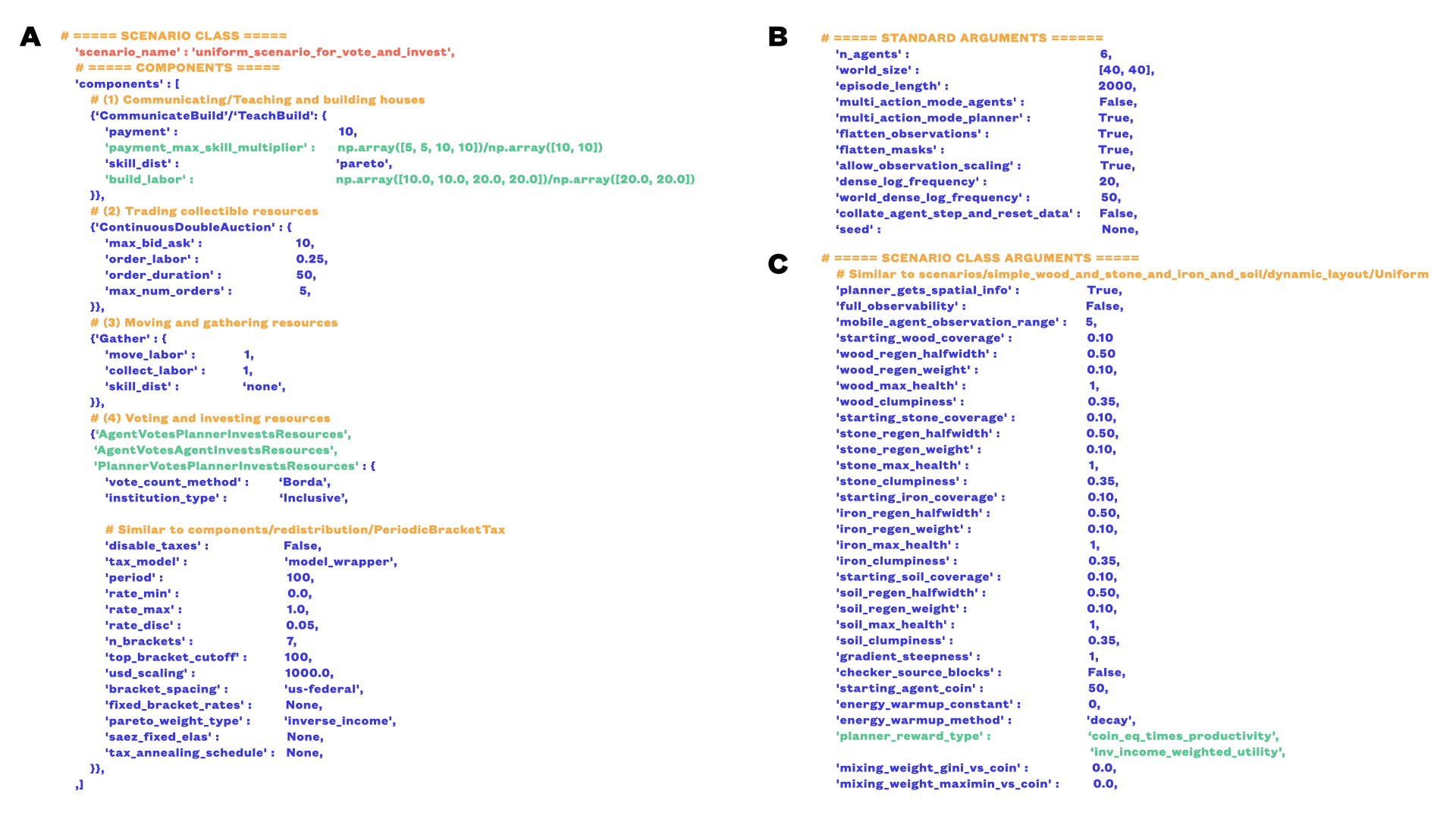}
	\caption{(A)(B)(C) Different features and input parameters of the Modified AI-Economist with Communication and with Teaching which are tested and their aggregated plots are brought and discussed in the main text. The orange texts indicate various parts of the input structure. The green texts show the alternative parameters which are tested in this paper.}
	\label{Figure11}
\end{figure}

\newpage

\begin{figure}
	\centering
	\includegraphics[width=0.7\linewidth]{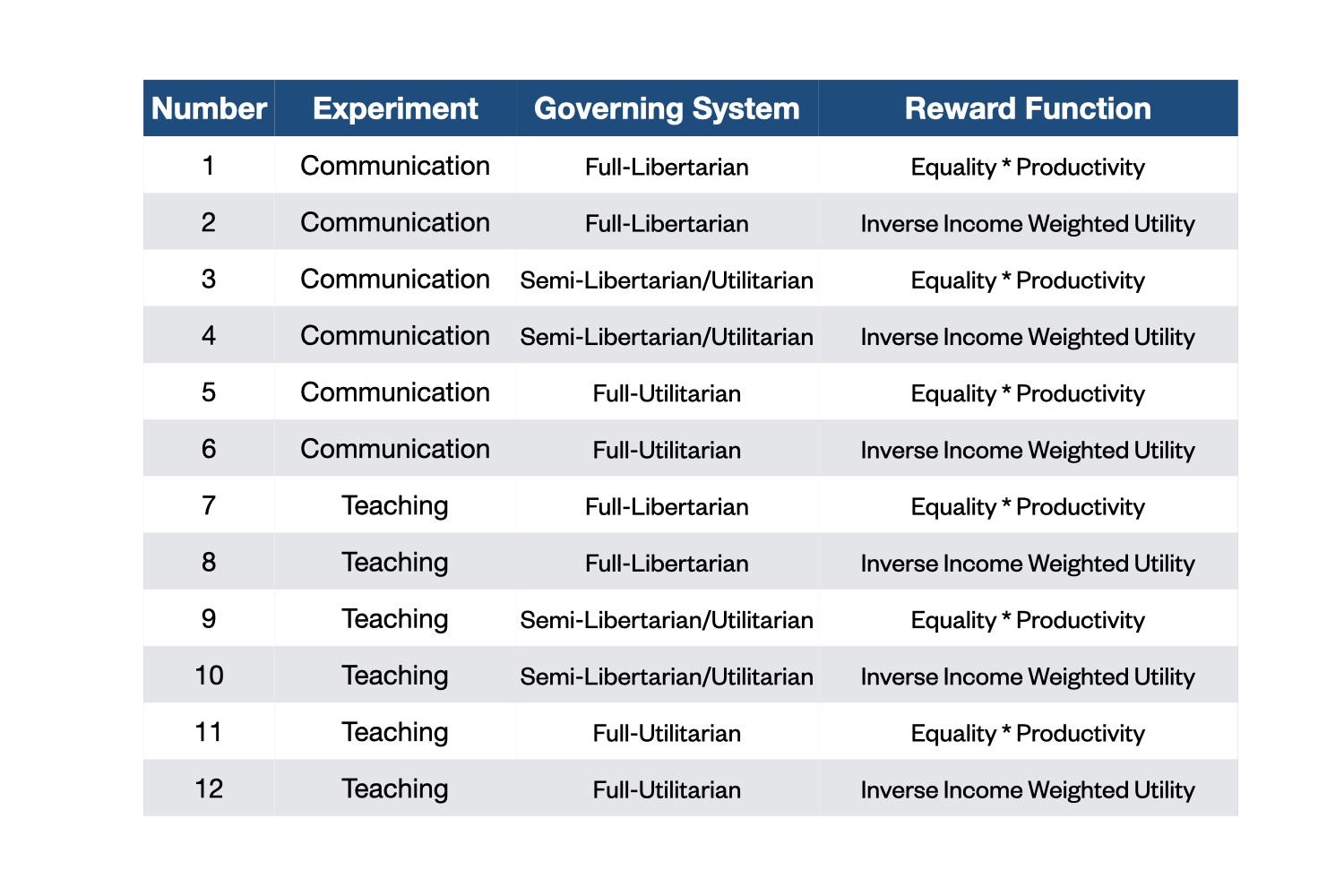}
	\caption{A figure showing all different runs of the Modified AI-Economist with Communication and with Teaching with different values as input parameters. The \textit{Reward Function} refers to the reward function of the central planner. To generate the plots in the main text, the generated results of a pair of consecutive simulations belonging to one kind of experiment and one governing system are pooled together.}
	\label{Figure12}
\end{figure}

\newpage

\begin{figure}
	\centering
	\includegraphics[width=0.7\linewidth]{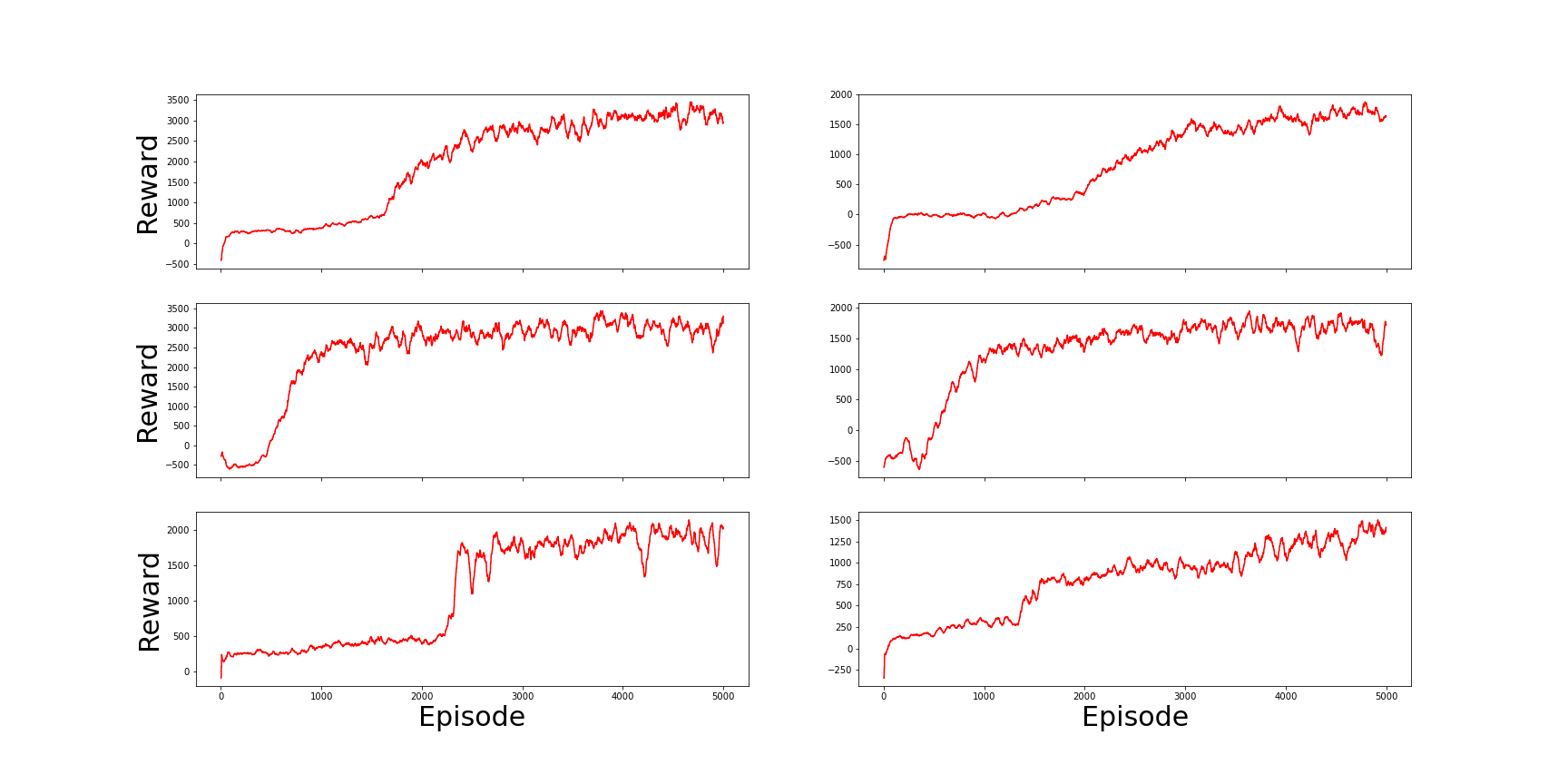}
	\caption{Average episode reward across training - 5000 episodes - for all runs of the Modified AI-Economist with Communication. The plots of the 6 runs 1-6 of Fig.~\ref{Figure12} are brought in the order from left-to-right and top-to-bottom. It is worthwhile to mention that the training of two-level RL is particularly unstable, but it seems that almost all the simulations have been converged, but they are less stable than the plots of Fig.~\ref{Figure14}}
	\label{Figure13}
\end{figure}

\newpage

\begin{figure}
	\centering
	\includegraphics[width=0.7\linewidth]{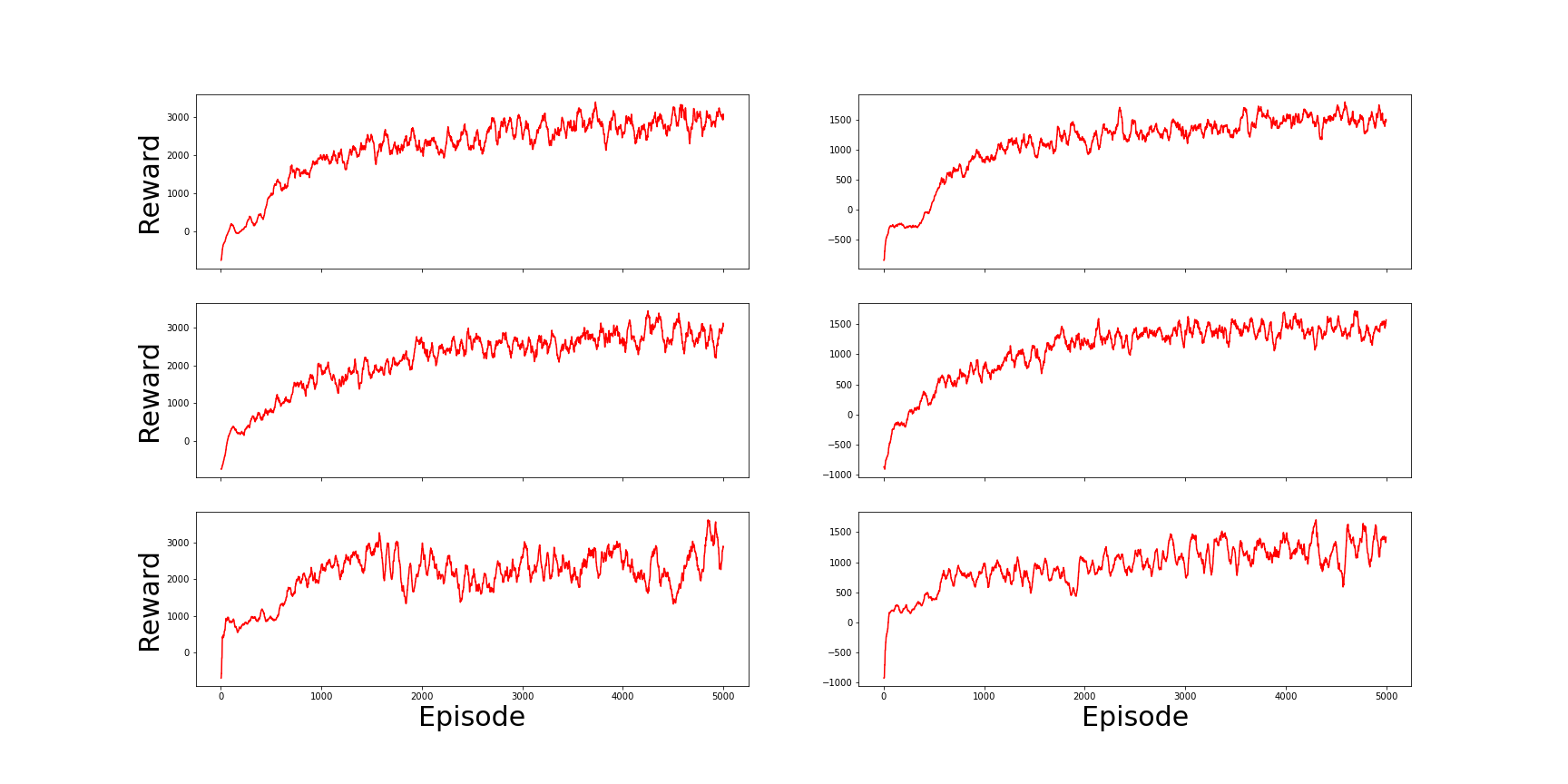}
	\caption{Average episode reward across training - 5000 episodes - for all runs of the Modified AI-Economist with Teaching. The plots of 6 runs 7-12 of Fig.~\ref{Figure12} are brought in the order from left-to-right and top-to-bottom. It is worthwhile to mention that the training of two-level RL is particularly unstable, but it seems that almost all the simulations have been converged, and they are more stable than the plots of Fig.~\ref{Figure13}}
	\label{Figure14}
\end{figure}

\end{document}